\documentclass[a4paper]{article}

\usepackage{INTERSPEECH2022}

\usepackage{multirow,booktabs}
\usepackage{caption}
\usepackage{subcaption}

\title{The Vicomtech Spoofing-Aware Biometric System for the SASV Challenge}
\name{Juan M. Mart\'in-Do\~{n}as, Iv\'an G. Torre, Aitor \'Alvarez and Joaquin Arellano}
\address{
  Vicomtech Foundation, Basque Research and Technology Alliance (BRTA),\\ Mikeletegi 57, 20009 Donostia -- San Sebasti\'an (Spain)}
\email{\{jmmartin, igonzalez, aalvarez, jarellano\}@vicomtech.org}

\begin{document}

\maketitle
\begin{abstract}
This paper describes our proposed integration system for the spoofing-aware speaker verification challenge. It consists of a robust spoofing-aware verification system that use the speaker verification and antispoofing embeddings extracted from specialized neural networks. First, an integration network, fed with the test utterance's speaker verification and spoofing embeddings, is used to compute a spoof-based score. This score is then linearly combined with the cosine similarity between the speaker verification embeddings from the enrollment and test utterances, thus obtaining the final scoring decision. Moreover, the integration network is trained using a one-class loss function to discriminate between target trials and unauthorized accesses. Our proposed system is evaluated in the ASVspoof19 database, exhibiting competitive performance compared to other integration approaches. In addition, we test, along with our integration approach, state of the art speaker verification and antispoofing systems based on self-supervised learning, yielding high-performance speech biometric systems.
\end{abstract}


\section{Introduction}

Recent advances in deep neural network (DNN) architectures have led to significant improvements in the performance of speaker verification (SV) systems \cite{snyder2018x, desplanques2020ecapa, mary2022}, broadening its application in real scenarios as access controls or assistant devices. However, these biometric systems can be vulnerable to spoofing attacks \cite{tan21}, which try to access the system by replicating the target speaker's voice using different strategies such as voice synthesis and conversion, replay attacks, or even adversarial examples \cite{kreuk2018fooling}. Therefore, countermeasure (CM) systems \cite{wu15, nautsch2021} have been developed to detect and reject these attacks successfully. Robust deep learning-based approaches have been continuously integrated to deal with the increasingly complex spoofing attacks scenarios \cite{monteiro2020, jung2021aasist, martin2022vicomtech}.

Despite the efforts on developing single systems for each task, the integration of complete biometric security systems (verification and antispoofing) is still undergoing in the research community. Integrated systems usually suffer significant performance degradation, especially when operating in tandem. Few previous works have explored the joint optimization of these systems, with strategies from the integration of both models \cite{todisco2018,gomez2021,zhang2022} to the develop of joint architectures \cite{sizov2015,li2020,shim2020}.

The Spoofing Aware Speaker Verification (SASV) challenge 2022 \cite{jung2022sasv} aims to improve State of the Art (SOTA) robustness to both zero-effort impostor access attempts and spoofing attacks. The SASV challenge focuses on evaluating the performance of integrated systems where both CM and SV subsystems are optimized together to improve the reliability of the complete system in both scenarios. Thus, the goal is to assess the performance of a joined system discriminating between different target speaker access and unauthorized accesses. This closer to reality scenario is much more challenging and less explored than dealing with isolated cases of one type of attack. Only data from ASVspoof 2019 \cite{wang2020asvspoof} and VoxCeleb 2 \cite{chung2018voxceleb2} datasets can be used for training and testing the systems, while evaluations are performed on the ASVspoof 2019 dataset.

Two baseline systems were provided at the beginning of the challenge based on the same pre-trained SV and CM subsystems. \textit{Baseline1} sums the scores produced by the subsystems, yielding poor performance. On the other hand, \textit{Baseline2} is more elaborated, relaying on a three hidden layer fully connected neural network feed with three embeddings: two of them extracted from the speaker verification system using the enrolment and test utterances. In contrast, the third one is obtained from the antispoofing model using the test utterance.

In this work, we present an integrated system based on pre-trained SV and antispoofing models. First, a feed-forward vanilla neural network was trained to compute a spoofing score from previously extracted SV and spoofing embeddings of the test utterance. Then, the final SASV score is obtained from the SV and antispoofing ones using a linear operation. The integration model was trained using a one-class loss function focused on target trials, computing higher scores. We evaluated our approach using the challenge's base SV and spoofing systems, reporting a $0.84\%$ error rate for the SASV challenge. Moreover, we also evaluated more advanced SV and CM systems from the ones of the challenge, including self-supervised learning approaches based on wav2vec2 \cite{baevski20}, yielding a competitive biometric system with a $0.14$\% error rate.

The remainder of this paper is organized as follows. First, in Section \ref{sec:system}, we detail the proposed spoofing-aware integration system. Then, in Section \ref{sec:results} the experimental framework and discussed results are presented. Finally, main highlights and conclusions are discussed in Section \ref{sec:conclusions}.

\section{Proposed system}
\label{sec:system}

Our approach is based on an integrated SASV system that combines a neural network-based spoofing metric followed by a linear combination of SV and spoofing scores. Let us first consider two base systems, trained for antispoofing and SV, which compute a single embedding per utterance. Thus, we define $\mathbf{y}_{\textrm{sv}}$ as the SV embedding of the enrollment utterance, while $\mathbf{x}_{\textrm{sv}}$ and $\mathbf{x}_{\textrm{spf}}$ are the SV and spoofing embeddings from the test utterance, respectively. The SV score is directly obtained as the cosine similarity between the SV embeddings as $S_{\textrm{sv}} = \cos(\mathbf{y}_{\textrm{sv}}, \mathbf{x}_{\textrm{sv}}) \in [-1,1]$.

Our objective is to compute a spoofing score for the test utterance that considers both embeddings from the base systems. To this end, we propose an integration network fed with the concatenation of both test embeddings: $\mathbf{x}_{\textrm{sv}}$ and $\mathbf{x}_{\textrm{spf}}$. This network is similar to the \textit{Baseline2} system of the SASV challenge, and it includes three feed-forward layers with LeakyReLU activations and $256$, $128$, and $64$ hidden units, respectively. A batch normalization layer is added at the input of the deep neural network (DNN) to improve convergence during training by regularizing the variance of the embedding units. Moreover, a linear layer is placed after the feed-forward layers to compute a new 64-dimension spoofing embedding $\mathbf{e}_{\textrm{spf}}$. The output of the network is a spoofing score computed as the cosine similarity $S_{\textrm{spf}} = \cos(\mathbf{w}, \mathbf{e}_{\textrm{spf}})$, where $\mathbf{w}$ is a trained vector network parameter representing the direction of genuine speech in the embedding space.

Finally, the SASV score can be obtained as a linear combination of the previously computed scores as follows,
\begin{equation}
S_{\textrm{sasv}} = \alpha S_{\textrm{sv}} + S_{\textrm{spf}},
\end{equation}
where $\alpha$ is a scalar value optimized during the network training phase. The integration network is trained to obtain high $S_{\textrm{sasv}}$ for target genuine trials. Inspired by previous works \cite{zhang2021,mingote2021}, the proposed integration network is trained by using a one-class softmax loss function. Given a batch of $N$ trials, the loss is computed as
\begin{equation}
\mathcal{L}_{\textrm{OCS}} = \frac{1}{N} \sum_{n=1}^{N} \log\left( 1 + e^{\beta \left(m_{z_n} - S_{\textrm{sasv,n}} \right)\left( -1 \right)^{z_n}} \right),
\end{equation}
where $n$ is the batch trial index, $z$ is $0$ for the target class and $1$ otherwise (non-target and spoof classes), $\beta$ is a scale factor, and $m_z$ is a class-depending margin. Different margins allow broadening the expected scores for the attack classes (impostors and spoofing), thus increasing the robustness.

Our proposal hypothesizes that a better spoofing score can be obtained when considering both the information contained in the SV and the antispoofing test embeddings. Preliminary experiments regarding other approaches --including only the spoofing embedding or the three different embeddings-- yielded higher errors, supporting our previous theses. Furthermore, while the use of SV test embedding allows more robust decisions, the enrollment SV embedding degraded the performance of the integration system. On the other hand, the SV score is explicitly considered in the final score computation. This strategy lets the integration system focus on challenging spoofing attacks (those that result in high SV scores). At the same time, the SV information helps to discriminate them from zero-effort impostors and weaker attacks for the SV system. This strategy is also explored in \cite{zhang2022}, but instead of following a probabilistic framework, we linearly combine the SV and spoofing scores, obtaining better results.

It is worthwhile to notice a relevant feature of our proposed system with respect to other integration approaches, such as the one followed in \textit{Baseline2}: the modularity of the SV system. Our system does not require the enrollment embedding to be processed by the integration network because the SV score is directly used in the final SASV score. Therefore, this system is compatible with different \textit{homomorphic encryption} schemes \cite{nautsch2019}, that allow certain operations as cosine similarity or linear score combination in the encrypted domain. Thus, the enrollment embeddings can be encrypted in the biometric system database, preventing unauthorized users from accessing private biometric information.

\section{Experimental results}
\label{sec:results}

\begin{table*}[!t] 
\renewcommand{\arraystretch}{1.3}
\caption{EER (\%) results of our proposed integration system on ASVspoof19, both development and evaluation sets. The results for the baseline systems and other approaches are also shown for comparison purposes.}
\label{table:sasv_results}
\centering
\begin{tabular}{lcccccc}
\toprule
\multicolumn{1}{c}{\multirow{2}{*}{\textbf{System}}} & \multicolumn{3}{c}{\textbf{Development}} & \multicolumn{3}{c}{\textbf{Evaluation}} \\
\multicolumn{1}{c}{}                  & SV-EER & SPF-EER & SASV-EER & SV-EER & SPF-EER & SASV-EER      \\ \midrule
ECAPA (SV) \cite{desplanques2020ecapa}     & 1.86 & 20.28 & 17.37    & 1.64  & 30.76 & 23.84 \\
Cascade CM-SV         & 1.89 & 0.42  & 1.08     & 1.64  & 6.59  & 4.44  \\
Logistic regression     & 2.70 & 0.67  & 1.68     & 2.55  & 2.52  & 2.55  \\
Baseline2 \cite{jung2022sasv}       & 9.58 & 0.12  & 4.04     & 11.29 & 0.65  & 6.24  \\
Zhang et al. \cite{zhang2022} & 2.02 & \textbf{0.07} & 1.10  & 1.94  & 0.80  & 1.53  \\
\midrule
\textbf{Proposed system}  & \textbf{1.08} & 0.13 & \textbf{0.54}      
                          & \textbf{0.97} & \textbf{0.58} & \textbf{0.84} \\
\bottomrule
\end{tabular}
\end{table*}

This section first describes the experimental framework, including the database used for training and evaluating our system, the trainable hyperparameters, and the evaluation metrics. Then, we reveal the results obtained for the SASV challenge 2022 and some additional evaluation scores using more discriminative SV and CM subsystems.

\subsection{Experimental framework}

Our proposed SASV system was evaluated in the logistic access (LA) partition of the ASVspoof19 \cite{wang2020asvspoof} database. The database is derived from the VCTK corpus \cite{yamagishi12}, and it includes both bonafide speech and spoofing utterances generated by using different speech synthesis and voice conversion algorithms. The database is split into training, development, and evaluation subsets with non-overlapped speakers. Both development and evaluation subsets include protocols for evaluating automatic speaker verification systems with spoofing attacks. Therefore, there are three different kinds of trials: target --bonafide test utterance from the same speaker as the enrollment one--, non-target --bonafide test utterance from an impostor speaker--, and spoof --synthetic test utterance--.

The challenge baselines are based on the AASIST system \cite{jung2021aasist}, trained on the LA train partition of the ASVspoof2019 for CM, and the ECAPA-TDNN model \cite{desplanques2020ecapa} trained on VoxCeleb 2 \cite{chung2018voxceleb2} dataset, for SV. AASSIST integrates a RawNet2-based encoder \cite{tak2021end} to extract high-level representations from raw waveform inputs in order to feed a graph attention network \cite{velivckovic2017graph} used for the extraction of CM embeddings. Meanwhile, ECAPA-TDNN is based upon Res2Net architecture \cite{gao2019res2net} with a squeeze-excitation module to model channel interdependencies \cite{hu2018squeeze}.

The different approaches were evaluated in terms of three different equal error rate (EER) metrics, according to the kind of trials compared. These metrics are the SV-EER --target vs. non-target trials--, the SPF-EER --target vs. spoof trials--, and the SASV-EER --target vs. non-target and spoof trials--. While SASV-EER gives the total performance of the integrated system, the other two metrics offer a more granular detail of the system behavior.

The parameters selected for the loss function were $\beta=20$, $m_0=0.9$ and $m_1=0.2$. The model was trained using the Adam optimizer \cite{kingma15} with a learning rate of $1e^{-4}$ and a mini-batch of $24$ trials. The training was run during $40$ epochs, keeping the model with the best SASV-EER in the development set for evaluation.

\subsection{Challenge results}

\begin{figure}[!t]
 \centering
 \begin{subfigure}{\linewidth}
     \centering
     \includegraphics[width=\linewidth]{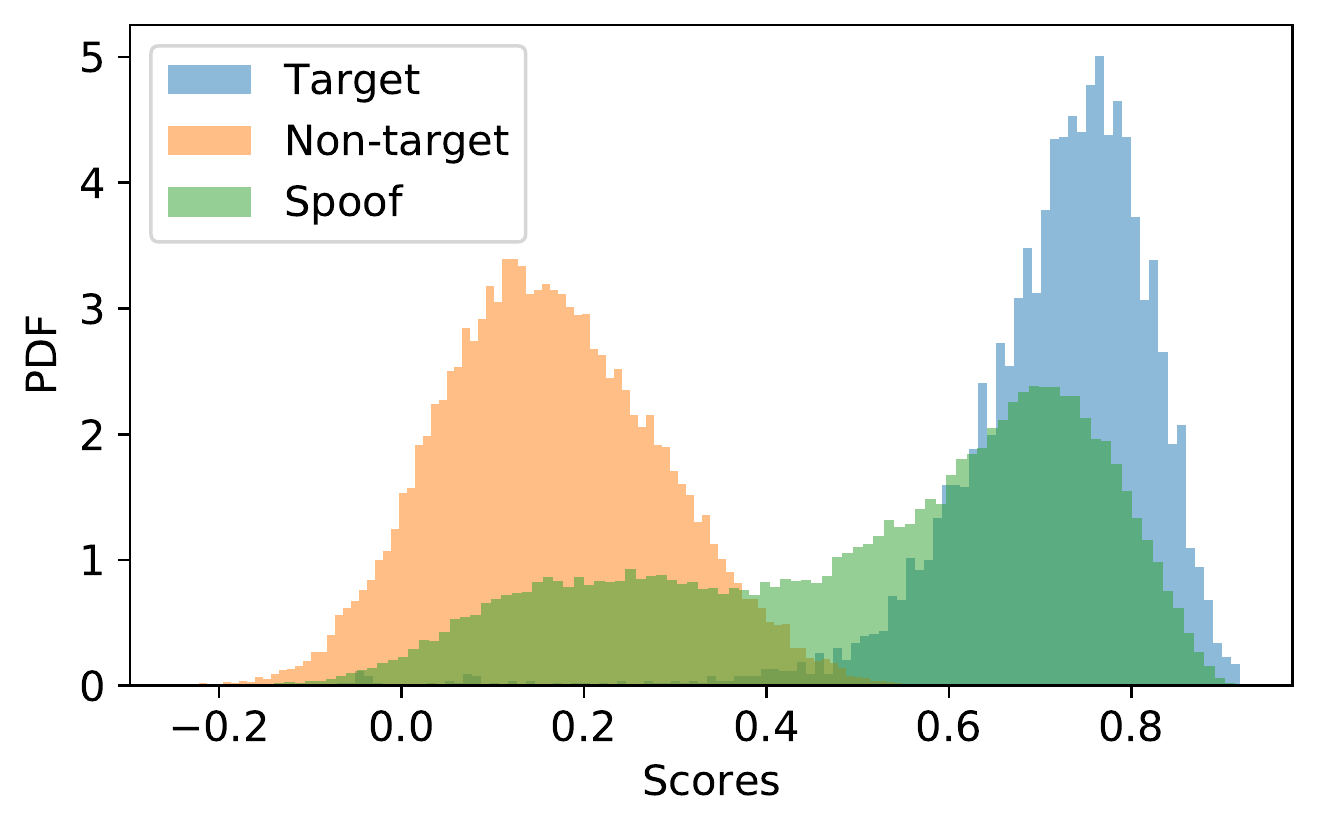}
     \caption{ECAPA-TDNN}
 \end{subfigure}
 \begin{subfigure}{\linewidth}
     \centering
     \includegraphics[width=\linewidth]{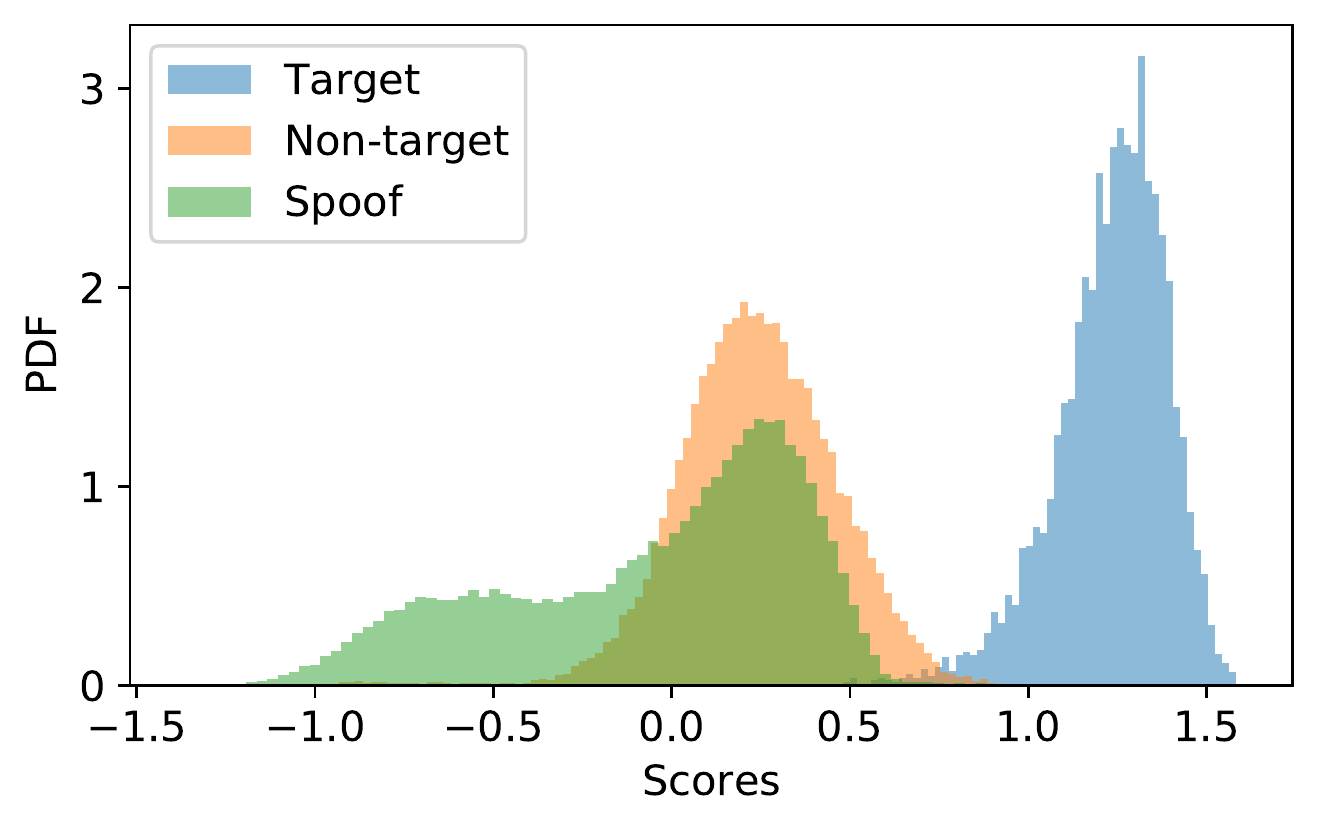}
     \caption{Proposed system using ECAPA and AASIST}
     \label{fig:three sin x}
 \end{subfigure}
 \caption{
  \label{fig:hist_base}%
 Comparison between score empirical probability distribution functions for the ECAPA-TDNN verification system (top) and our proposed integration system (below). Performance scores are obtained by testing the evaluation set of ASVspoof19.}
\end{figure}

In this subsection, we compared our proposed system with the \textit{Baseline2} integrated network and the probabilistic fusion framework proposed in \cite{zhang2022}. Moreover, we also evaluated two more integration approaches:

\begin{itemize}
\item A cascade CM-SV approach that first detects spoofing utterances --spoofing score under a given threshold-- and then computes the SV score. In this configuration, the threshold was selected by minimizing the error in the development set.

\item A logistic regression approach that combines base systems SV and spoofing scores to compute a SASV score as the probability of being a target trial. The logistic regression was trained using the scores from the development set trials.
\end{itemize}
All the evaluated integration systems are based on ECAPA-TDNN and AASIST for SV and antispoofing tasks, respectively, using the allowed data according to the challenge rules.

Table~\ref{table:sasv_results} shows the EER obtained for the three evaluated tasks on the development and evaluation subsets of ASVspoof19. Our proposal outperforms the other systems, achieving the lowest EER in the three different evaluations tasks and two tasks of the development set. Overall, the results on the evaluation set achieve a $0.84\%$ of SASV-EER, a $0.58\%$ of SPF-EER, and a $0.97\%$ of SV-EER. The different spoofing-aware approaches perform better than the SV ECAPA baseline but with different behaviors. The \textit{Baseline2} obtains competitive results for spoofing --$0.65\%$ of SPF-EER-- but degrades the SV performance --$11.29\%$ of SV-EER--. Both cascade and logistic regression approaches were optimized in the development set, obtaining good results --$1.08\%$ and $1.68\%$ SASV-EER,
respectively--, but significantly degraded in the evaluation set --$4.44\%$ and $2.55\%$ SASV-EER respectively--. This is particularly significant for the cascade system, where the spoofing threshold may not be optimal on the evaluation set. The proposed approach in \cite{zhang2022} obtains competitive results on all evaluation tasks --$1.94\%$, $0.80\%$ and $1.53\%$ for SV, SPF, and SASV, respectively--. Nevertheless, the SV and antispoofing performance are still lower than the ECAPA model in SV-EER. Our proposed system outperforms previously published methods and reduces the SV-EER of the ECAPA system while keeping spoofing detection performance similar to \textit{Baseline2} approach. These results demonstrate that our proposal can effectively exploit the information in the test embeddings along the SV scoring to obtain a competitive SASV system.

In addition, \figurename{~\ref{fig:hist_base}} shows the score empirical probability distribution functions for the ECAPA-TDNN (top) compared with our proposed integration system (below). As it can be observed, there is a high overlap between the target and spoof score distributions in the ECAPA-TDNN, revealing that it is not robust against spoofing attacks. On the other hand, our proposed system achieves better separation between target scores and the remaining classes, translating to lower EERs, as can be seen in Table \ref{table:sasv_results}.

\subsection{Evaluation using SOTA SV and CM subsystems}

\begin{table*}[!t] 
\renewcommand{\arraystretch}{1.3}
\caption{EER (\%) results for evaluating our proposed integration system using different SV and CM subsystems. Results for the ECAPA2 and W2V2 base SV subsystems are included for comparison purposes.}
\label{table:sota_results}
\centering
\begin{tabular}{cccccccc}
\toprule
\multicolumn{2}{c}{\textbf{Subsystem}} & \multicolumn{3}{c}{\textbf{Development}} & \multicolumn{3}{c}{\textbf{Evaluation}} \\
SV & CM & SV-EER & SPF-EER & SASV-EER & SV-EER & SPF-EER & SASV-EER      \\ \midrule
ECAPA2 & -  & 1.21 & 17.86 & 15.23    & 0.76 & 27.01 & 20.63 \\
W2V2-SV & -  & 0.81 & 14.80 & 12.39    & 0.50 & 20.42 & 15.73  \\
\midrule
ECAPA & AASIST   & 1.08 & 0.13 & 0.54    & 0.97 & 0.58 & 0.84 \\
ECAPA2 & AASIST  & 0.88 & 0.20 & 0.46    & 0.45 & 0.58 & 0.52  \\
ECAPA2 & W2V2-CM    & 0.68 & \textbf{0.00} & 0.33    & 0.49 & 0.26 & 0.35  \\
W2V2-SV & AASIST    & 0.73 & 0.07 & 0.40    & 0.45 & 0.46 & 0.45 \\
W2V2-SV & W2V2-CM      & \textbf{0.40} & \textbf{0.00} & \textbf{0.13}    & \textbf{0.18} & \textbf{0.11} & \textbf{0.14}  \\
\bottomrule
\end{tabular}
\end{table*}

In order to further evaluate the benefits of our proposed system, in this subsection, we evaluated our approach using different SOTA base SV and CM subsystems. For speaker verification, two additional networks were considered:
\begin{itemize}

\item ECAPA2: The ECAPA-TDNN previously evaluated but, in this case, trained using speech data from both Voxceleb 1 and 2 datasets.

\item W2V2-SV: This method is based on the self-supervised learning paradigm, using a wav2vec2 (W2V2) architecture \cite{baevski20}. We evaluated the pipeline system proposed in \cite{chen21}, which combines W2V2 with ECAPA-TDNN downstream model, with the models finetuned along using Voxceleb 2 data.

\end{itemize}
Regarding the antispoofing task, we included our previously proposed approach based on W2V2 (W2V2-CM) \cite{martin2022vicomtech}. This model comprises a pre-trained wav2vec2 feature extractor and a spoofing network classifier trained on ASVspoof19. We chose the best model for our evaluations in ASVspoof 2021 LA \cite{yamagishi21}, which used the cross-lingual W2V2 model pre-trained with 128 languages and narrowband FIR filtering data augmentation during model finetuning. The spoofing embedding was selected from the attentive pooling layer instead of the final linear layer, as these embeddings showed better robustness in our preliminary experiments. Moreover, the SV and spoofing embeddings of the evaluated subsystems were first length-normalized before feeding them to the integration network, thus avoiding norm scale issues.

\begin{figure}[!t]
  \centering
  \includegraphics[width=\linewidth]{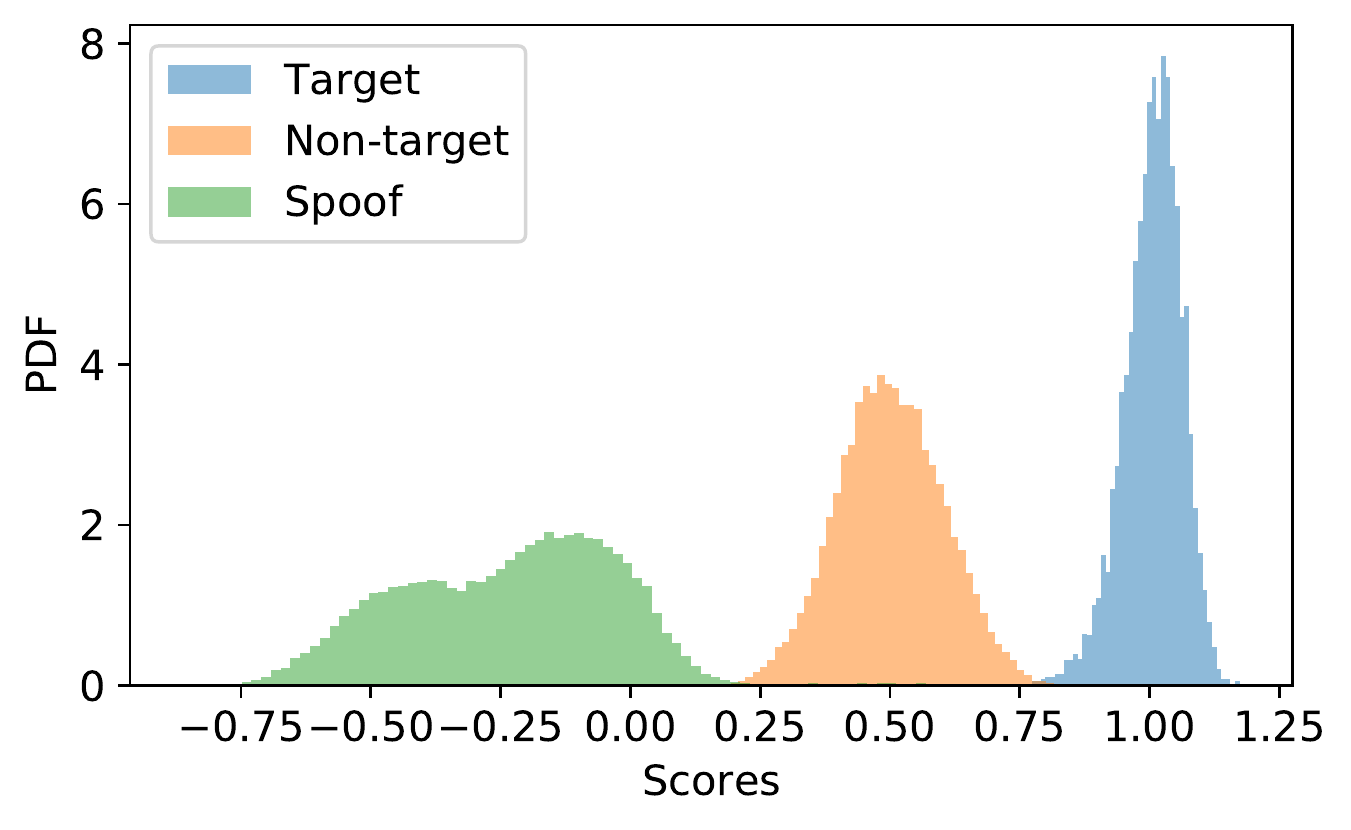}
  \caption{Score empirical probability distribution functions of the integrated system using wav2vec2 architectures for both SV and CM subsystems. The performance is computed by testing the evaluation set of ASVspoof19.}
  \label{fig:hist_sasv_w2v2}
\end{figure}

Table~\ref{table:sasv_results} shows the different EER obtained when evaluating several combinations of these base subsystems with our proposed integration approach. The results for the ECAPA2 and W2V2-SV systems (without using spoofing countermeasures) are also shown for comparison purposes. The two considered SV subsystems perform better than the ECAPA baseline in SV and antispoofing subtasks. Moreover, the integration approach with these systems and W2V2-CM further increases the performance. In particular, the W2V2-based subsystems outperform the remaining ones, with the combination of both W2V2 models yielding a SASV-EER of 0.14\% in the evaluation set. This is a very competitive result comparable with the best approach presented in the SASV challenge\footnote{https://sasv-challenge.github.io/challenge\_results/}, which obtained a 0.13\% SASV-EER. The winner system is based on robust SV and CM integrated systems, scoring normalization, and an additional ensemble classifier. On the other hand, we perform comparable with a more straightforward integration network approach by relying on high-performance W2V2 subsystems.

Finally, \figurename{~\ref{fig:hist_sasv_w2v2}} shows the empirical probability distribution function of scores obtained for our proposed integrated approach along with the W2V2-based subsystems. This system gets a significantly better separation between the three different classes, with the target trials achieving the highest scores and the spoofing trials the lowest scores.

\section{Conclusions}
\label{sec:conclusions}

In this work, we have presented our proposed spoofing-aware speaker verification system for the SASV challenge. Our proposed integration network exploited the SV and spoofing embeddings of the test utterance to compute a robust spoofing score. The final SASV score was obtained as a linear combination of the SV and spoofing scores. Moreover, the model was trained to compute higher scores for target trials using a one-class softmax loss function. Our proposed method showed good results in the ASVspoof19 development and evaluation sets, outperforming other spoofing-aware approaches and performing significantly better than the base systems in the SV and spoofing detection tasks. Furthermore, the performance increased when combining our integration approach with robust W2V2-based SV and CM subsystems, yielding competitive results among the top systems in the SASV challenge. As future work, we will evaluate the finetuning of the complete SASV system and different training strategies.

\section{Acknowledgements}

This work was supported in part by the Spanish Centre for the Development of Industrial Technology (CDTI) through the Project ÉGIDA—RED DE EXCELENCIA EN TECNOLOGIAS DE SEGURIDAD Y PRIVACIDAD under Grant CER20191012.

\bibliographystyle{IEEEtran}

\bibliography{mybib}

\end{document}